\documentclass{article}
    \PassOptionsToPackage{numbers, compress}{natbib}

 \usepackage[main, final]{neurips_2026}

\usepackage[utf8]{inputenc} 
\usepackage[T1]{fontenc}    
\usepackage{hyperref}       
\usepackage{url}            
\usepackage{booktabs}       
\usepackage{amsfonts}       
\usepackage{nicefrac}       
\usepackage{microtype}      
\usepackage{xcolor}         
\usepackage{amsmath}
\usepackage{algorithm, algorithmic}
\usepackage{enumitem} 
\usepackage{graphicx}
\usepackage{multirow}

\title{TwiSTAR:Think Fast, Think Slow, Then Act,\\Generative Recommendation with Adaptive Reasoning}

\author{
    \textbf{Shiteng Cao$^{\spadesuit}$} \quad
    \textbf{Kaian Jiang$^{\spadesuit}$} \quad
    \textbf{Yunlong Gong$^{\spadesuit}$} \quad
    \textbf{Zhiheng Li$^{\spadesuit}$} \\
    $^{\spadesuit}$Shenzhen International Graduate School, Tsinghua University \\
    \textnormal{\{caost24, jka23, gongyl25\}@mails.tsinghua.edu.cn,\ zhhli@tsinghua.edu.cn}
}

\begin{document}

\maketitle

\begin{abstract}
Generative recommendation with Semantic IDs (SIDs) has emerged as a promising paradigm, yet existing methods apply a \emph{fixed} inference strategy, either fast direct generation or slow chain-of-thought reasoning, uniformly across all user histories. This approach creates a trade-off: fast recommendation model produces suboptimal accuracy on hard samples, while always invoking slow reasoning incurs prohibitive latency and wastes computation on easy cases. To address this, we propose \textbf{TwiSTAR}, a framework that learns to \emph{adaptively allocate reasoning effort} per user sequence. Our system equips an LLM with three complementary tools: a fast SID-based retriever, a lightweight candidate ranker, and a slow reasoning model that generates explicit rationales before recommending. Crucially, we inject collaborative commonsense into the slow model by transforming item-to-item knowledge into natural language explanations. A planner, trained through supervised warm-up followed by agentic reinforcement learning, dynamically decides which tool to invoke. Experiments on three datasets demonstrate that our method outperforms strong baselines, achieving consistent accuracy gains while reducing inference latency compared to uniform slow reasoning. 

\end{abstract}

\section{Introduction}

Recommender systems play a fundamental role in mitigating information overload, connecting users with relevant items in domains ranging from e-commerce to content streaming. Their effectiveness directly impacts user satisfaction and business metrics. In recent years, generative recommendation has emerged as a powerful paradigm, framing sequential recommendation as an autoregressive generation task: given a user's interaction history, the model predicts the identifier of the next item. Among identifier choices, Semantic IDs (SIDs) obtained from residual quantization (e.g., RQ-VAE) have proven particularly attractive. SIDs form a lightweight vocabulary that naturally aligns with the token space of large language models (LLMs). This approach enables seamless integration with LLMs and achieves strong performance by leveraging both collaborative signals and language priors. However, existing methods suffer from a fundamental limitation: they apply a \emph{uniform inference strategy} to all user histories, regardless of their complexity or ambiguity.

Two predominant inference modes exist. \textit{Fast thinking} directly generates the target SID without explicit intermediate reasoning. It is computationally efficient and well-suited for routine patterns, but often fails on long-tail scenarios~\citep{10.1145/3626772.3657690}. \textit{Slow reasoning} first produces a chain-of-thought (CoT) rationale, then outputs the SID, improving accuracy on hard cases and enhancing interpretability~\citep{wei2023chain}. 
OneRec-Think~\citep{liu2025onerecthinkintextreasoninggenerative} adopts a ``Think-Ahead'' architecture for industrial deployment: the computationally expensive reasoning chain and the first two itemic tokens are generated offline and cached, while an online model completes the generation under prefix constraints. This decoupling ensures real-time responsiveness but still applies the same fixed two-stage reasoning to every request, lacking adaptive control.
GREAM~\citep{hong2025generativereasoningrecommendationllms} supports direct sequence recommendation and sequential reasoning recommendation, but the mode must be pre-specified per request; the system cannot dynamically choose which mode to use based on the user's history.
OxygenRec~\citep{hao2025oxygenrec} introduces a fast-slow architecture, where a near-line LLM synthesizes contextual instructions and an efficient online backbone performs low-latency generation; however, the decision to invoke reasoning remains heuristic rather than learned. Consequently, an effective and efficient fusion of fast and slow reasoning, where the system adaptively allocates reasoning effort based on the difficulty of each user history, remains an open challenge.
Despite recent progress in reasoning-enhanced recommenders~\citep{liu2025onerecthinkintextreasoninggenerative, hong2025generativereasoningrecommendationllms, hao2025oxygenrec}, none adaptively allocate reasoning effort based on query difficulty.

\begin{figure*}[t]
  \centering
  \includegraphics[width=0.9\textwidth]{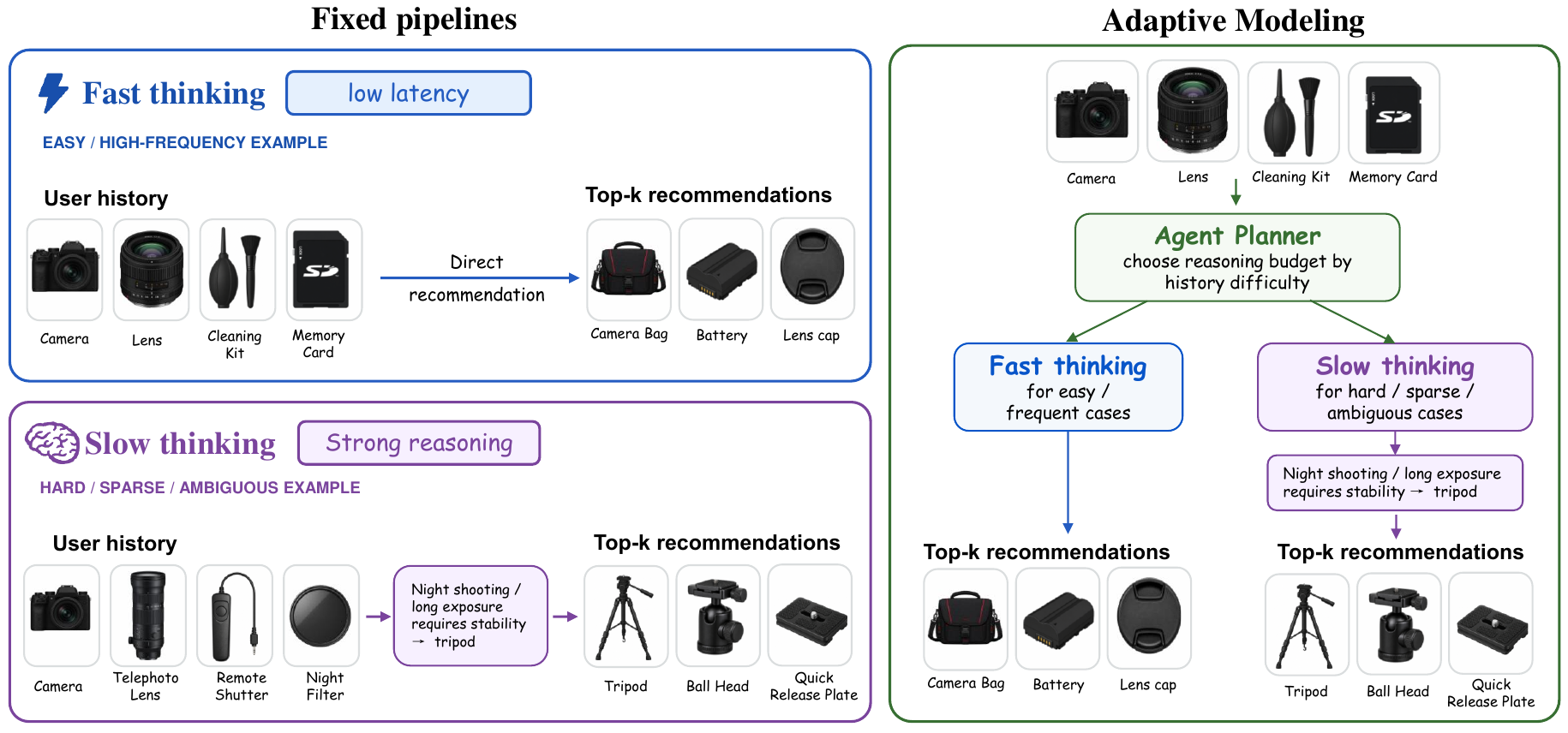} 
\caption{Illustration of fast vs. slow reasoning in generative recommendation. Left: fixed separate pipelines without adaptive selection. Right: our proposed framework with an agent planner that conditionally invokes fast retriever or slow reasoning model.
}
  \label{fig:intro}
\end{figure*}

This paper addresses a central question: \textit{Can a generative recommender learn when to think fast and when to think slow?} Recent LLMs have begun to address a similar trade-off by moving from single-mode instruction following toward adaptive inference-time computation~\citep{qwen2025qwen3,microsoft2025phi4reasoning,anthropic2025extendedthinking}. Inspired by the dual-system view of cognition, modern models increasingly support efficient responses for simple queries and deliberate reasoning for complex ones.
As Figure~\ref{fig:intro} shows, we propose \textbf{TwiSTAR}, \textbf{T}wo‑mode thinking, \textbf{S}low and fast \textbf{T}hinking, then \textbf{A}ct with tools to \textbf{Recommend}, a generative recommendation framework that learns to adaptively allocate reasoning effort. Our system is built upon a base model that aligns SID space with the LLM's text embedding space. This alignment grounds SID tokens in the textual contexts of their corresponding items, serving as the foundation for all subsequent components.
Specifically, we develop three tools. 
A fast recommendation model directly predicts SIDs via efficient top-\(k\) retrieval, offering low-latency inference. A slow reasoning model first generates an explicit Chain-of-Thought rationale before outputting an SID; to instill genuine collaborative commonsense, we extract I2I relations from historical co-occurrence patterns, convert them into natural language explanation instructions, and activate the reasoning capability via RL. For cases where fast retrieval alone is insufficient but full slow reasoning is unnecessarily expensive, we introduce a ranking model, which retrieves a larger candidate set with the fast recommendation model and then applies a discriminative ranker to reorder the results. Finally, an agent planner decides, for each user sequence, which tool to invoke, enabling adaptive reasoning allocation tailored to the difficulty of the history.

Our main contributions are:
\begin{itemize}[leftmargin=*]
    \item A method to enhance collaborative reasoning in LLM-based recommenders by transforming I2I relations into natural language explanations.
    \item A two-stage training recipe (supervised imitation + agentic RL) that teaches the planner to selectively invoke slow reasoning, which unifies fast SID-based generation, slow reasoning, and an adaptive planner into a single agentic system. To our knowledge, this is the first work to learn when to engage reasoning in generative recommendation.
    \item Extensive experiments on three public datasets show that our approach consistently outperforms strong baselines in both ranking accuracy and efficiency.
\end{itemize}


\section{Related Work}
\label{sec:related}
\subsection{LLM as Recommender System}
\label{sec:rw_llm4rec}
LLMs provide a natural interface for recommendation by representing users, items, and preferences in text. A straightforward direction encodes items with structured templates, thereby formulating recommendation within the native vocabulary space of LLMs. Prior text-based LLM recommenders instantiate this idea through product-description generation~\citep{cui2022m6recgenerativepretrainedlanguage}, instruction-style prompting~\citep{10.1145/3604915.3608857}, hierarchical attribute modeling~\citep{zhang2024llmtreerecunleashingpowerlarge}, preference-optimized generation~\citep{10.1145/3716393,10.5555/3737916.3738779}, multimodal item summarization~\citep{10.1007/978-981-97-2650-9_3}, and language-processing formulations that combine textual features with behavioral signals~\citep{geng2023recommendationlanguageprocessingrlp,chu2023leveraging,10.1145/3626772.3657690}. Despite their flexibility and interpretability, purely text-based representations are inefficient for large-scale recommendation and may fail to ground generated text to a unique catalog item.
Semantic-ID-based generative recommendation addresses this limitation by representing each item as a compact sequence of discrete tokens and casting recommendation as catalog-grounded sequence generation. LC-Rec~\citep{10597986} highlights the gap between language semantics in LLMs and collaborative semantics. PLUM~\citep{10.1145/3774904.3792802} adapts pretrained LLMs to industrial-scale generative recommendation through continued pre-training and task-specific fine-tuning. 
TS-Rec~\citep{feng2026finegrainedsemanticsintegrationlarge} investigates token-level SID semantics by initializing SID token embeddings with semantic-aware signals. These studies show that SID tokens should be semantically and collaboratively aligned rather than treated as opaque symbols.
Recent work further introduces explicit reasoning into generative recommendation. OneRec-Think~\citep{liu2025onerecthinkintextreasoninggenerative} and SIDReasoner~\citep{he2026reasoningsemanticidsenhances} extend generative recommendation to reasoning through itemic alignment and reasoning activation. Other industrial systems~\citep{li2025leadremultifacetedknowledgeenhanced,damico2026deployingsemanticidbasedgenerative,denadai2026unified, liang2026generativereasoningreranker} demonstrate that LLM-based recommenders are advancing toward reasoning-capable and production-scale deployment.
 Despite this progress, existing models lack adaptive reasoning allocation, applying the same inference strategy regardless of query difficulty. Consequently, they cannot balance efficiency and effectiveness across user histories of varying complexity. More severely, the abundance of easy samples that fast recommendation model can already handle tends to degrade the performance of slow reasoning models, leading to undesirable collapse.

\subsection{Agent for Recommender System}
\label{sec:agent_for_recommender}
Recent agentic recommender systems explore different aspects of process-level decision making. AMEM4Rec~\citep{nguyen2026amem4rec} introduces evolving memory for agentic LLM recommenders, capturing collaborative signals through cross-user memory evolution. RecGPT-V2~\citep{yi2025recgptv2} develops a hierarchical multi-agent framework for industrial user-intent reasoning and multi-objective optimization. RecBot~\citep{tang2025recbot} and TalkPlay~\citep{doh2025talkplay} further investigate interactive recommendation agents that translate natural language requests into actionable strategies or tool calls. Together, these works show that LLM recommenders are moving beyond passive item generation toward agents with adaptive reasoning, memory, interaction, and tool-use capabilities.
However, existing agentic recommenders typically optimize one aspect of the recommendation process. They do not fully address how a SID-based generative recommender should jointly allocate inference behaviors under heterogeneous user histories.  
We formulate generative recommendation as an agentic inference problem: the model should decide not only \emph{what} to recommend, but also \emph{how} to recommend, whether to think fast, invoke a ranking tool, or think slow with collaborative reasoning.

\section{Preliminaries}
\label{sec:prelim}

We first introduce Semantic IDs (SIDs) and formulate sequential recommendation as autoregressive generation over SIDs.

Given an item \(i \in \mathcal{I}\) with textual description \(t_i\), we encode it into a continuous embedding:
\begin{equation}
\mathbf{e}_i = \mathrm{Encoder}(t_i).
\end{equation}
We then apply residual k-means to quantize \(\mathbf{e}_i\) into a sequence of discrete codes. Let \(L\) denote the number of quantization layers and \(K\) the codebook size per layer. Starting from \(\mathbf{r}_{i,0} = \mathbf{e}_i\), the quantization at layer \(j\) is given by
\begin{equation}
c_{i,j} = \arg\min_{k \in \{1,\ldots,K\}} 
\left\| \mathbf{r}_{i,j-1} - \mathbf{v}_{j,k} \right\|_2^2,
\qquad
\mathbf{r}_{i,j} = \mathbf{r}_{i,j-1} - \mathbf{v}_{j,c_{i,j}},
\label{eq:residual_kmeans_prelim}
\end{equation}
where \(\mathbf{v}_{j,k}\) is the \(k\)-th codeword in the \(j\)-th codebook. The resulting Semantic ID of item \(i\) is
\begin{equation}
\mathrm{SID}(i) = [c_{i,1}, c_{i,2}, \ldots, c_{i,L}].
\label{eq:sid_def_prelim}
\end{equation}

For a user \(u\), let \(\mathcal{H}_u = [i_1, i_2, \ldots, i_{T-1}]\) denote the historical interaction sequence, and let \(i_T\) be the next item to predict. Replacing each item with its SID, sequential recommendation is formulated as autoregressive SID prediction:
\begin{equation}
p\!\big(\mathrm{SID}(i_T) \mid \mathrm{SID}(i_{<T})\big)
=
\prod_{j=1}^{L}
p\!\big(c_{i_T,j} \mid c_{i_T,<j}, \mathrm{SID}(i_{<T})\big),
\label{eq:sid_autoregressive_prelim}
\end{equation}
where \(c_{i_T,<j} = [c_{i_T,1}, \ldots, c_{i_T,j-1}]\). This formulation allows a pretrained language model to perform recommendation via constrained token generation over valid SIDs.
 
\section{Methodology}
\label{sec:method}
\begin{figure*}[t]
  \centering
  \includegraphics[width=\textwidth]{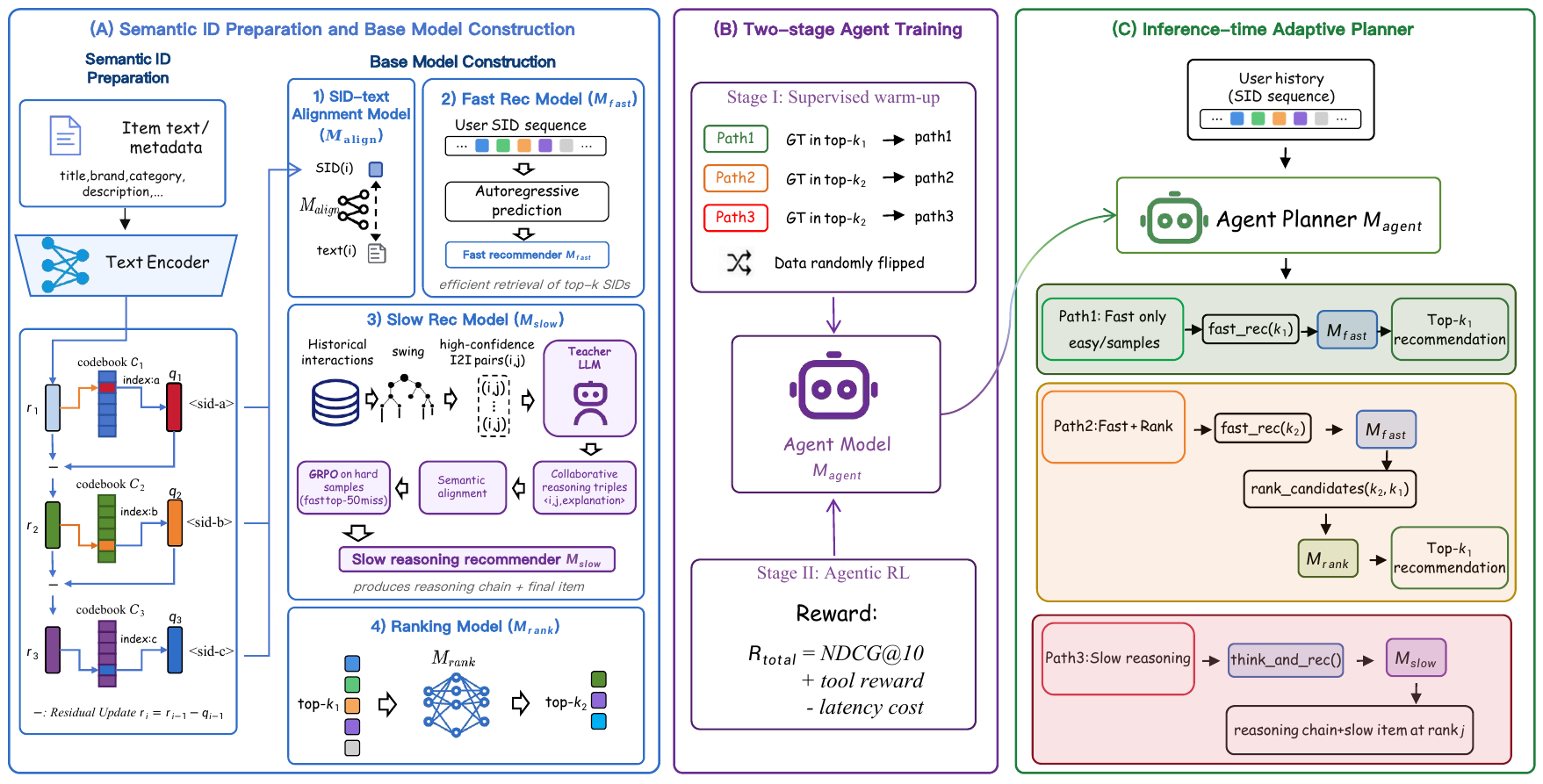} 
    \caption{Our framework first extracts SIDs from item metadata and aligns them with text embeddings to ground semantic meaning. It then trains a fast SID-based retrieval model. Next, it injects collaborative commonsense into the slow recommendation model and activates reasoning via reinforcement learning. In the second stage, the planner is trained through supervised warm-up followed by reinforcement learning to balance accuracy and latency. This agentic design ensures that reasoning is applied only when beneficial, avoiding the inefficiency of uniform slow reasoning.}
  \label{fig:overview}
\end{figure*}

To overcome the limitations of uniform inference in generative recommendation, as illustrated in Figure~\ref{fig:overview}, we propose a two-stage framework that learns to adaptively allocate reasoning effort per user sequence. The core idea is to equip a semantically aligned language model with three specialized tools: a fast SID-based retriever for routine patterns, a ranking model for candidate refinement, and a slow chain-of-thought model for hard cases that require explicit reasoning. A learned agent planner then dynamically decides which tool to invoke.

\subsection{Aligned Base Model}
\label{sec:alignment}

The core challenge of using SIDs in a generative recommender is that initial SID tokens, which are obtained via residual k-means (Section~\ref{sec:prelim}), are discrete codes without inherent linguistic meaning. A pretrained language model cannot directly interpret these tokens, which hinders effective next-item prediction. To bridge this gap, we align SID token embeddings with the natural language semantics of their corresponding items, while keeping the main parameters of the LLM fixed.

Specifically, for each item \(i\), we construct an item-level alignment sequence by pairing its SID with its textual metadata, such as its title, category, and other available metadata.
We feed these item-level sequences into a pretrained causal language model and optimize the standard language modeling objective:
\begin{equation}
\mathcal{L}_{\text{align}}
= - \sum_{i \in \mathcal{I}} \sum_{m=1}^{|x_i|}
\log p_{\theta}(x_{i,m} \mid x_{i,<m}).
\label{eq:align_loss}
\end{equation}

After alignment, SID tokens are grounded in the textual contexts of their corresponding items. As a result, when an SID appears in later recommendation sequences, the LLM can process it as a semantically meaningful item representation rather than an opaque discrete code. We denote the resulting model as \(\mathcal{M}_{\text{align}}\), which serves as the foundation for the subsequent fast, slow, and agentic components.

\subsection{Fast Rec Model and Ranking Model}

After alignment, we train \(\mathcal{M}_{\text{fast}}\) to perform sequential recommendation. Given a user’s interaction history represented by their SIDs, the model autoregressively predicts the next item’s SID:
\begin{equation}
\mathcal{L}_{\text{rec}}
= -\sum_{u}\sum_{t}\sum_{\ell=1}^{L}
\log p_{\mathcal{M}_{\text{fast}}}
(c_{t,\ell}\mid c_{t,<\ell}, \text{SID}(i_{<t})).
\end{equation}

Training data is constructed offline by truncating each user’s chronological sequence into input–target pairs, where the input contains only the SID tokens of historical items. During inference, we constrain decoding to valid SIDs via a prefix trie and retrieve top‑\(K\) candidates with beam search. This yields a fast recommendation model that retrieves candidate items efficiently.

The ranking model \(\mathcal{M}_{\text{rank}}\) is built upon the Deep Interest Network \citep{zhou2018din}, which captures the relevance between a user's historical behavior sequence and a candidate item through a local activation unit. Training data is constructed as follows: for each user, we retrieve \(K_2\) candidates using \(\mathcal{M}_{\text{fast}}\). If the ground truth is not among them, we add it as a positive sample and randomly sample \(K_1-1\) negatives from the retrieved set. We monitor AUC on a validation set for early stopping.

\subsection{Slow Rec Model: Collaborative Reasoning as Language Injection}
The core idea is to transform implicit collaborative signals into explicit natural language reasoning chains, thereby teaching the model \textit{what} co-occurrence patterns are and \textit{why} they exist.
Directly eliciting explanations is noisy and prone to hallucination: the true next item is only one of many plausible continuations, and the model often overfits to sequential id correlations without capturing robust statistical regularity. In contrast, I2I relations aggregate over many user sequences, providing denser, more reliable statistical signals. Learning from I2I explanations forces the model to internalize stable co‑occurrence patterns, rather than memorizing fragile sequential shortcuts. For each related pair \((i, j)\), we generate a reasoning instruction: ``In collaborative filtering, item \(i\) and item \(j\) are highly correlated. Please explain why users who purchase \(i\) also tend to purchase \(j\)?''

A teacher LLM (e.g., Qwen3.5-397B-A17B~\cite{qwen3.5}) produces a diverse set of explanations, forming triples \(\langle i, j, \text{explanation}\rangle\). These triples are then mixed with semantic alignment data to fine-tune the aligned recommendation model.  
Formally, let \(\mathcal{D}_{\text{collab}}\) be the set of (instruction, explanation) pairs derived from I2I relations. The training loss is a standard language modeling loss on \(\mathcal{D}_{\text{collab}}\). 
After this stage, the model acquires \textit{collaborative commonsense}: it does not memorize the I2I table but learns to verbalize statistical co-occurrence logic, enabling zero-shot reasoning on unseen pairs. We denote this semantically aligned and reasoning‑enhanced model as \(\mathcal{M}_{\text{base}}\), with all parameters fixed in the subsequent routing learning phase.

To further elicit reasoning before recommendation, we train a \textit{think} model using GRPO. Only those user sequences where the fast recommendation model fails to hit the ground truth within its HitRate@50 predictions are selected as training data, this focuses the reasoning ability on long‑tail and difficult samples.
The detailed specifications of reward function are provided in Appendix~\ref{app:reward}. The resulting model \(\mathcal{M}_{\text{slow}}\) can produce explicit chain‑of‑thought reasoning before outputting a recommendation.

\subsection{Planner Agent: Two‑Stage Agent Training}
\label{planner_agent}
We freeze \(\mathcal{M}_{\text{fast}}\) and \(\mathcal{M}_{\text{slow}}\) and optimize an agentic model \(\mathcal{M}_{\text{agent}}\) using a two‑stage procedure. The agentic model is initialized from the aligned base model; it takes a user’s sequence as input and outputs tool calls. Tools include:

\begin{itemize}[leftmargin=*]
    \item \texttt{fast\_rec(k)}: calls \(\mathcal{M}_{\text{fast}}\) to beam search k SIDs.
     \item \texttt{rank\_candidates(m, n)}: invokes the ranking model \(\mathcal{M}_{\text{rank}}\), which takes \(m\) items as input and outputs the top-\(n\) items after reordering.
    \item \texttt{think\_and\_rec(j)}: invokes \(\mathcal{M}_{\text{slow}}\) to produce a reasoning chain and then beam search \(j\) SIDs as the final recommendation.
\end{itemize}

We generate pseudo‑labels for the agent by evaluating \(\mathcal{M}_{\text{fast}}\) on a held‑out set:

\begin{itemize}[leftmargin=*]
    \item Path 1 (fast only): If the ground truth is already retrieved by the fast recommendation model with a small candidate set size \(K_1\), the agent invokes the fast recommendation model with that \(K_1\).
    \item Path 2 (fast + ranking): For each user, we retrieve \(K_2\) candidates using \(\mathcal{M}_{\text{fast}}\). If the ground-truth item is included in the retrieved set, it is treated as the positive item. If the ground truth is absent, we insert it as the positive item and sample negatives from the retrieved candidates.
    \item Path 3 (slow reasoning): \(\mathcal{M}_{\text{slow}}\) generates \(K_1\) candidate SIDs through constrained beam search, and these candidates are used as the final top-\(K_1\) list.
\end{itemize}

To encourage exploration during supervised warm-up, we randomly relabel a small fraction of Path-1 samples as higher-cost paths. The agentic model is trained with a standard cross‑entropy loss on the sequence of tool calls and arguments.

After supervised warm‑up, we further refine \(\mathcal{M}_{\text{agent}}\) using reinforcement learning (e.g., GRPO or PPO). The reward is designed to balance final recommendation quality and process efficiency:
\begin{equation}
R_{\text{total}} = \underbrace{\text{NDCG@10}(\text{final ranked list})}_{\text{outcome reward}} \;+\; \underbrace{\eta \cdot \mathbb{I}[\text{valid tool sequence}]}_{\text{process reward}} \;-\; \beta \cdot \text{(latency cost)}.
\end{equation}

\section{Experiment}
\label{sec:experiments}
\textbf{Datasets.}
We conduct experiments on three real-world recommendation datasets from the Amazon review benchmark~\cite{mcauley2015imagebasedrecommendationsstylessubstitutes}: \textsc{Beauty}, \textsc{Toys and Games} and \textsc{Sports and Outdoors}. For evaluation, we adopt the leave one out strategy: the last interaction of each user is held out for testing, the second-last for validation, and all preceding interactions for training. We evaluate over the full item catalog rather than sampled negatives.

\textbf{Baselines.}
We compare our proposed method against a diverse set of competitive baselines, covering traditional sequential recommenders, generative recommendation models, and reasoning-enhanced approaches. All baseline results are reproduced under the same experimental setting with consistent evaluation protocols. A detailed description of each baseline is provided in Appendix~\ref{app:baseline}.

\textbf{Implementation Details.}
We implement our framework using Qwen3.5-4B as the base LLM. Semantic IDs are generated with $L=3$ quantization layers and codebook size $K=256$ per layer, yielding a vocabulary of $3 \times 256 = 768$ SID tokens. For the fast recommendation model, we use beam search with width $k=50$. Due to space constraints, we report detailed implementations in Appendix~\ref{app:impl_details}.

\textbf{Evaluation Metrics.}
We adopt two standard ranking metrics: Recall@K and Normalized Discounted Cumulative Gain@K (NDCG@K). For each user, the model outputs a top-$K$ recommendation list $\mathcal{R}_u(K)$. Recall@K measures whether the ground-truth item $i_u$ appears in this list.
NDCG@K further accounts for the ranking position.
\section{Discussion}
In this section, we address our five research questions: RQ1 confirms the overall superiority of our framework across diverse datasets, RQ2 quantifies the individual contributions of tools, RQ3 validates that explicit I2I explanation injections instill genuine collaborative commonsense, RQ4 demonstrates that hard-sample training improves the usefulness of slow reasoning on difficult cases, and RQ5 reveals the planner's decisions substantially align with oracle routing patterns.
\subsection{Overall Performance (RQ1)}
\begin{table}[htbp]
\setlength{\tabcolsep}{4pt}
\centering
\caption{Overall performance comparison of our method and baselines on three datasets. The best results are in \textbf{bold} and the second-best results are \underline{underlined}.}
\label{tab:overall}
\begin{tabular}{lcccccccc}
\toprule
Dataset & Metric & HGN & GRU4Rec & SASRec & TIGER & HSTU & OneRec-Think & Ours \\
\midrule
\multirow{4}{*}{Beauty} & R@5 & 0.0325 & 0.0392 & 0.0397 & 0.0409 & 0.0418 & \underline{0.0557} & \textbf{0.0609} \\ 
& R@10 & 0.0531 & 0.0585 & 0.0606 & 0.0622 & 0.0648 & \underline{0.0770} & \textbf{0.0880} \\
& N@5 & 0.0197 & 0.0263 & 0.0258 & 0.0267 & 0.0280 & \underline{0.0390} & \textbf{0.0415} \\
& N@10 & 0.0267 & 0.0325 & 0.0320 & 0.0337 & 0.0352 & \underline{0.0461} & \textbf{0.0504} \\
\midrule
\multirow{4}{*}{Sports} & R@5 & 0.0188 & 0.0190 & 0.0199 & 0.0219 & 0.0263 & \underline{0.0281} & \textbf{0.0324} \\
& R@10 & 0.0316 & 0.0312 & 0.0306 & 0.0342 & 0.0348 & \underline{0.0401} & \textbf{0.0476} \\
& N@5 & 0.0114 & 0.0122 & 0.0107 & 0.0137 & 0.0168 & \underline{0.0188} & \textbf{0.0214} \\
& N@10 & 0.0155 & 0.0157 & 0.0146 & 0.0179 & 0.0221 & \underline{0.0228} & \textbf{0.0257} \\
\midrule
\multirow{4}{*}{Toys} & R@5 & 0.0327 & 0.0330 & 0.0447 & 0.0338 & 0.0365 & \underline{0.0553} & \textbf{0.0594} \\
& R@10 & 0.0522 & 0.0491 & 0.0621 & 0.0546 & 0.0561 & \underline{0.0774} & \textbf{0.0828} \\
& N@5 & 0.0193 & 0.0223 & 0.0300 & 0.0210 & 0.0244 & \underline{0.0389} & \textbf{0.0425} \\
& N@10 & 0.0255 & 0.0279 & 0.0357 & 0.0277 & 0.0308 & \underline{0.0461} & \textbf{0.0505} \\
\bottomrule
\end{tabular}
\end{table}

Table~\ref{tab:overall} reports the overall performance comparison between various baseline models and our proposed method on three Amazon review datasets. Across all datasets and metrics, our method consistently outperforms all baselines. 
Among the baselines, OneRec-Think shows the strongest overall performance and consistently ranks second, although it consumes more computational resources during both training and inference. These results demonstrate the effectiveness of our proposed approach in capturing user preferences and generating accurate recommendations across diverse product categories.

\subsection{Ablation Study (RQ2)}
\label{sec:ablation}
To quantify the contribution of each component, we compare the full model against five variants on the Beauty dataset: (1) \textbf{w/o Ranking} (ranker removed); (2) \textbf{w/o Slow Reasoning} (slow model removed); (3) \textbf{w/o Slow Reasoning and Ranking} (both removed); (4) \textbf{w/o Alignment, Slow Reasoning and Ranking} (plus random SID initialization); (5) \textbf{Slow Reasoning Only w/o Planner} (slow model applied to every sequence).


\begin{figure}[htbp]
  \centering
  \includegraphics[width=0.75\columnwidth]{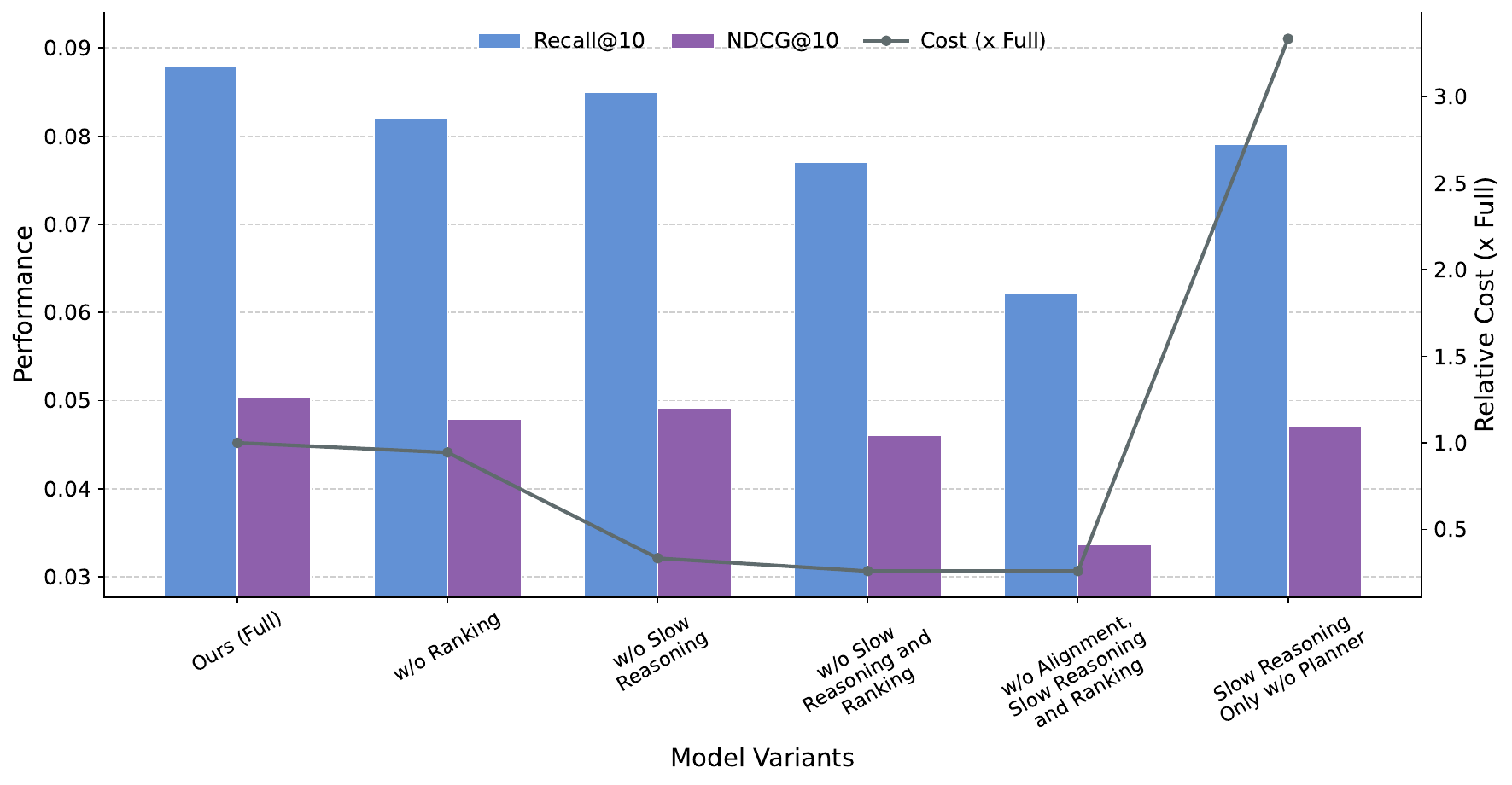} 
  \caption{Recall@10, NDCG@10 and relative inference cost (normalized to the full model) of different ablation variants on the Beauty dataset. The full model achieves the best performance while maintaining moderate cost. Removing any component degrades accuracy, and applying slow reasoning uniformly (without planner) incurs more time cost with lower accuracy.}
  \label{fig:ablation}
\end{figure}


Removing the ranking model (w/o Ranking) degrades R@10 from 0.0880 to 0.0820 and N@10 from 0.0504 to 0.0479, while only slightly reducing cost (0.94×). This indicates that the ranking tool provides a non‑trivial accuracy boost by refining the fast recommendation model’s candidate set, yet its own overhead is modest. Removing slow reasoning (w/o slow reasoning) causes a smaller drop, but its cost reduces dramatically. This shows that the slow model is expensive but brings substantial gains when invoked selectively. When both slow reasoning and ranking are removed (w/o slow reasoning and Ranking), performance falls further, confirming that each tool contributes uniquely to the final accuracy.
Aligning SID tokens with natural language semantics is the bedrock of our framework. Without alignment, even the fast recommendation model’s retrieval becomes severely impaired, and the reasoning capability cannot be activated.
The slow reasoning Only variant, which applies the slow reasoning model uniformly without the planner, achieves lower performance than the full model while incurring more cost. (All reported cost values are normalized to the full model’s total inference latency, which already includes the planner’s own decision‑making overhead.) 
This does not imply that slow reasoning is universally superior. Since \(\mathcal{M}_{\text{slow}}\) is intentionally specialized on hard samples, applying it to easy histories may introduce unnecessary generation noise and degrade direct collaborative matching.
Therefore, the full model with the planner delivers better accuracy with much lower average latency than always using slow reasoning, even after accounting for the planner’s extra computation.
Absolute latency and throughput numbers (in milliseconds and queries per second) along with a detailed breakdown of the planner's overhead are reported in Appendix~\ref{app:latency}. In summary, the ablation study validates that all components are essential. Moreover, the adaptive planner is critical for translating the raw capability of slow reasoning into an efficient and effective system.

\subsection{Does Explicit I2I Explanation Work? (RQ3)}

A core claim of our method is that transforming I2I relations into natural language explanations and fine-tuning the LLM on these explanations injects genuine collaborative commonsense, rather than memorizing co‑occurrence tables. To test this claim, we design a discriminative evaluation task that directly measures whether the model has internalized the underlying collaborative similarity structure.
We first extract all unique I2I pairs from the Beauty dataset, resulting in 25,438 pairs \((i, j)\) where item \(j\) is frequently co‑purchased after item \(i\). 
The first-level SID codes of \(i\) and \(j\) are identical for 26.25\% of the I2I pairs. When both the first and second codes match, this occurs for only 3.65\% of the pairs.
For each such pair, we randomly sample 9 distractor items from the item catalog and construct a candidate set of 10 items: the true associated item \(j\) plus 9 negatives. The model is then presented with item \(i\) (described by its SID and optionally its textual metadata) and the list of 10 candidate SIDs. It must select the candidate that is most collaboratively similar to item \(i\). We report the accuracy of selecting the ground‑truth \(j\).  

We evaluate two model variants: the aligned model \(\mathcal{M}_{\text{align}}\) (without explicit I2I explanation training) and the slow recommendation model \(\mathcal{M}_{\text{slow}}\) (after fine‑tuning on I2I explanation instructions). The base model achieves only 49\% accuracy, indicating that alignment alone does not capture fine‑grained collaborative relationships. After injecting I2I explanations, accuracy jumps to 84\%. These results suggest that I2I explanation tuning helps the model better capture collaborative associations beyond the aligned SID representation.

\subsection{Why Does Thinking Need Hard Samples? (RQ4)}

We first train the slow reasoning model on the \emph{entire} training set. 
Under this setting, the think model becomes \emph{underperforming}: it fails to handle ambiguous requests better than the fast non‑think model. 
Specifically, we define \emph{easy samples} as those whose history contains less than two second‑level category (about 30\% of the entire dataset), and \emph{hard samples} as those with three or more distinct second‑level categories.
As shown in Table~\ref{tab:full_training_mediocre}, on the hard test set, the think model trained on full data is even slightly \emph{lower} than the fast recommendation model. 
This confirms that mixing easy and hard samples during training leads the model to fail to develop genuine reasoning ability for difficult cases.

In contrast, selective hard-sample training yields a modest but consistent improvement on hard cases, suggesting that reasoning supervision is more useful when concentrated on samples where direct generation fails.
A concrete case study is provided in Appendix~\ref{app:case_study_slow_rec_model} to illustrate this phenomenon.

\begin{table}[htbp]
\centering
\caption{Think model trained on full data or selective samples (Beauty dataset).}
\label{tab:full_training_mediocre}
\begin{tabular}{lcc}
\toprule
Model & Hard samples (R@10) & Easy Samples (R@10) \\
\midrule
Non‑think (fast) & 0.0706 & 0.0760 \\
Think (trained on full data) & 0.0692 & \textbf{0.0782} \\
Think (trained on hard samples) & \textbf{0.0723} & 0.0735 \\
\bottomrule
\end{tabular}
\end{table}

\subsection{Planner Decision Analysis (RQ5)}
\label{sec:planner_analysis}
We analyze the decisions made by the learned planner \(\mathcal{M}_{\text{agent}}\) on the Beauty test set. The planner is given a user history and must choose among three tools: (1) fast retrieval only (\texttt{fast\_rec}), (2) fast retrieval plus ranking (\texttt{fast\_rec+rank}), and (3) slow reasoning (\texttt{think\_and\_rec}). The planner invokes \texttt{fast\_rec} only for 62.3\% of sequences, \texttt{fast\_rec+rank} for 27.1\%, and \texttt{think\_and\_rec} for 10.6\%.
To understand whether these decisions align with difficulty, we compare the planner’s choices with an oracle that knows the performance of each tool. For each sequence, we compute the optimal tool as the one that yields the highest NDCG@10 (ties broken by lowest latency). The planner achieves 78.3\% agreement with the oracle. The most common mismatch is choosing \texttt{fast\_rec} when the oracle would prefer \texttt{fast\_rec+rank} (13.4\% of all sequences), i.e., regret due to underestimating the need for ranking. Conversely, the planner rarely chooses slow reasoning when fast only would suffice (1.7\%). The planner successfully identifies difficult sequences: among the 10.6\% where slow reasoning is invoked, the ground truth is not in the fast recommendation model’s top‑50 for 86.2\% of them, confirming that the planner targets genuine hard cases. Compared to a heuristic baseline that triggers slow reasoning when the number of distinct categories in the user's history exceeds a tuned threshold, our learned planner achieves higher accuracy (NDCG@10: 0.0504 vs. 0.0489), confirming the value of learning the routing policy.

\section{Conclusion and Future Work}
\label{sec:conclusion}

We presented TwiSTAR, a generative recommendation framework that learns to adaptively allocate reasoning effort. At its core, we unify a fast SID-based retriever, a ranking model, and a slow reasoning model that generates explicit rationales. By transforming I2I patterns into language explanations, the slow recommendation model achieves reasoning. A two‑stage training recipe, combining supervised imitation with reinforcement learning, teaches the planner when to invoke slow reasoning, when to rank, and when fast retrieval suffices. Extensive experiments on three Amazon review datasets demonstrate that our method consistently outperforms strong baselines, while ablation studies confirm the contribution of each component. Importantly, we show that the hard-sample training improves the usefulness of slow reasoning on difficult cases and that the planner’s decisions reduce average inference latency compared to uniform slow reasoning.
Beyond these empirical gains, our system offers a conceptual advance: it treats both the fast recommendation model (traditional recommendation systems) and the slow-thinking model (incorporates world knowledge and reasoning) as tools that an agent can execute. This design ensures seamless compatibility with existing recommendation approaches and opens up a novel paradigm for future recommendation systems, where heterogeneous reasoning modes are orchestrated by a learnable planner.

Despite these promising results, several limitations remain that point to directions for future work. First, our current implementation still incurs non‑negligible overhead for the slow reasoning model, even though it is invoked sparingly; future work could explore early termination of reasoning chains. Second, the planner’s decisions are learned offline; extending to online adaptation with bandit feedback could enable continuous improvement. Finally, while we focused on sequential recommendation, the idea of adaptive reasoning allocation may generalize to other generative tasks such as conversational recommendation or query rewriting, which we leave for future investigation.

\newpage
\bibliographystyle{unsrtnat}   
\bibliography{reference}

@misc{qwen3.5,
    title  = {{Qwen3.5}: Towards Native Multimodal Agents},
    author = {{Qwen Team}},
    month  = {February},
    year   = {2026},
    url    = {https://qwen.ai/blog?id=qwen3.5}
}

@misc{cui2022m6recgenerativepretrainedlanguage,
      title={{M6-Rec}: Generative Pretrained Language Models are Open-Ended Recommender Systems}, 
      author={Zeyu Cui and Jianxin Ma and Chang Zhou and Jingren Zhou and Hongxia Yang},
      year={2022},
      eprint={2205.08084},
      archivePrefix={arXiv},
      primaryClass={cs.IR},
      url={https://arxiv.org/abs/2205.08084}, 
}

@misc{geng2023recommendationlanguageprocessingrlp,
      title={Recommendation as Language Processing (RLP): A Unified Pretrain, Personalized Prompt {\&} Predict Paradigm (P5)}, 
      author={Shijie Geng and Shuchang Liu and Zuohui Fu and Yingqiang Ge and Yongfeng Zhang},
      year={2023},
      eprint={2203.13366},
      archivePrefix={arXiv},
      primaryClass={cs.IR},
      url={https://arxiv.org/abs/2203.13366}, 
}

@inproceedings{10.1145/3604915.3608857,
author = {Bao, Keqin and Zhang, Jizhi and Zhang, Yang and Wang, Wenjie and Feng, Fuli and He, Xiangnan},
title = {{TALLRec}: An Effective and Efficient Tuning Framework to Align Large Language Model with Recommendation},
year = {2023},
isbn = {9798400702419},
publisher = {Association for Computing Machinery},
address = {New York, NY, USA},
url = {https://doi.org/10.1145/3604915.3608857},
doi = {10.1145/3604915.3608857},
booktitle = {Proceedings of the 17th ACM Conference on Recommender Systems},
pages = {1007–1014},
numpages = {8},
location = {Singapore, Singapore},
series = {RecSys '23}
}

@misc{chu2023leveraging,
      title={Leveraging Large Language Models for Pre-trained Recommender Systems}, 
      author={Zhixuan Chu and Hongyan Hao and Xin Ouyang and Simeng Wang and Yan Wang and Yue Shen and Jinjie Gu and Qing Cui and Longfei Li and Siqiao Xue and James Y Zhang and Sheng Li},
      year={2023},
      eprint={2308.10837},
      archivePrefix={arXiv},
      primaryClass={cs.IR},
      url={https://arxiv.org/abs/2308.10837}, 
}

@article{10.1145/3716393,
author = {Bao, Keqin and Zhang, Jizhi and Wang, Wenjie and Zhang, Yang and Yang, Zhengyi and Luo, Yanchen and Chen, Chong and Feng, Fuli and Tian, Qi},
title = {A Bi-Step Grounding Paradigm for Large Language Models in Recommendation Systems},
year = {2025},
issue_date = {December 2025},
publisher = {Association for Computing Machinery},
address = {New York, NY, USA},
volume = {3},
number = {4},
url = {https://doi.org/10.1145/3716393},
doi = {10.1145/3716393},
journal = {ACM Trans. Recomm. Syst.},
month = apr,
articleno = {53},
numpages = {27}
}

@inproceedings{10.1007/978-981-97-2650-9_3,
author = {Karra, Saketh Reddy and Tulabandhula, Theja},
title = {{InteraRec}: Interactive Recommendations Using Multimodal Large Language Models},
year = {2024},
isbn = {978-981-97-2649-3},
publisher = {Springer-Verlag},
address = {Berlin, Heidelberg},
url = {https://doi.org/10.1007/978-981-97-2650-9_3},
doi = {10.1007/978-981-97-2650-9_3},
booktitle = {Trends and Applications in Knowledge Discovery and Data Mining: PAKDD 2024 Workshops, RAFDA and IWTA, Taipei, Taiwan, May 7–10, 2024, Proceedings},
pages = {32–43},
numpages = {12},
location = {Taipei, Taiwan}
}

@inproceedings{10.5555/3737916.3738779,
author = {Chen, Yuxin and Tan, Junfei and Zhang, An and Yang, Zhengyi and Sheng, Leheng and Zhang, Enzhi and Wang, Xiang and Chua, Tat-Seng},
title = {On softmax direct preference optimization for recommendation},
year = {2024},
isbn = {9798331314385},
publisher = {Curran Associates Inc.},
address = {Red Hook, NY, USA},
booktitle = {Proceedings of the 38th International Conference on Neural Information Processing Systems},
articleno = {863},
numpages = {27},
location = {Vancouver, BC, Canada},
series = {NIPS '24}
}

@inproceedings{10.1145/3626772.3657690,
author = {Liao, Jiayi and Li, Sihang and Yang, Zhengyi and Wu, Jiancan and Yuan, Yancheng and Wang, Xiang and He, Xiangnan},
title = {{LLaRA}: Large Language-Recommendation Assistant},
year = {2024},
isbn = {9798400704314},
publisher = {Association for Computing Machinery},
address = {New York, NY, USA},
url = {https://doi.org/10.1145/3626772.3657690},
doi = {10.1145/3626772.3657690},
booktitle = {Proceedings of the 47th International ACM SIGIR Conference on Research and Development in Information Retrieval},
pages = {1785–1795},
numpages = {11},
location = {Washington DC, USA},
series = {SIGIR '24}
}

@misc{zhang2024llmtreerecunleashingpowerlarge,
      title={{LLMTreeRec}: Unleashing the Power of Large Language Models for Cold-Start Recommendations}, 
      author={Wenlin Zhang and Chuhan Wu and Xiangyang Li and Yuhao Wang and Kuicai Dong and Yichao Wang and Xinyi Dai and Xiangyu Zhao and Huifeng Guo and Ruiming Tang},
      year={2024},
      eprint={2404.00702},
      archivePrefix={arXiv},
      primaryClass={cs.IR},
      url={https://arxiv.org/abs/2404.00702}, 
}

@INPROCEEDINGS{10597986,
  author={Zheng, Bowen and Hou, Yupeng and Lu, Hongyu and Chen, Yu and Zhao, Wayne Xin and Chen, Ming and Wen, Ji-Rong},
  booktitle={2024 IEEE 40th International Conference on Data Engineering (ICDE)}, 
  title={Adapting Large Language Models by Integrating Collaborative Semantics for Recommendation}, 
  year={2024},
  volume={},
  number={},
  pages={1435-1448},
  doi={10.1109/ICDE60146.2024.00118}}

@inproceedings{10.1145/3774904.3792802,
author = {He, Ruining and Heldt, Lukasz and Hong, Lichan and Keshavan, Raghunandan and Mao, Shifan and Mehta, Nikhil and Su, Zhengyang and Tsai, Alicia and Wang, Yueqi and Wang, Shao-Chuan and Yi, Xinyang and Baugher, Lexi and Cakici, Baykal and Chi, Ed and Goodrow, Cristos and Han, Ningren and Ma, He and Rosales, Romer and Soest, Abby Van and Tandon, Devansh and Wu, Su-Lin and Yang, Weilong and Zheng, Yilin},
title = {{PLUM}: Adapting Pre-trained Language Models for Industrial-scale Generative Recommendations},
year = {2026},
isbn = {9798400723070},
publisher = {Association for Computing Machinery},
address = {New York, NY, USA},
url = {https://doi.org/10.1145/3774904.3792802},
doi = {10.1145/3774904.3792802},
booktitle = {Proceedings of the ACM Web Conference 2026},
pages = {8093–8104},
numpages = {12},
location = {United Arab Emirates},
series = {WWW '26}
}

@misc{feng2026finegrainedsemanticsintegrationlarge,
      title={Fine-grained Semantics Integration for Large Language Model-based Recommendation}, 
      author={Jiawei Feng and Xiaoyu Kong and Leheng Sheng and Bin Wu and Chao Yi and Feifang Yang and Xiang-Rong Sheng and Han Zhu and Xiang Wang and Jiancan Wu and Xiangnan He},
      year={2026},
      eprint={2602.22632},
      archivePrefix={arXiv},
      primaryClass={cs.IR},
      url={https://arxiv.org/abs/2602.22632}, 
}

@misc{liu2025onerecthinkintextreasoninggenerative,
      title={{OneRec-Think}: In-Text Reasoning for Generative Recommendation}, 
      author={Zhanyu Liu and Shiyao Wang and Xingmei Wang and Rongzhou Zhang and Jiaxin Deng and Honghui Bao and Jinghao Zhang and Wuchao Li and Pengfei Zheng and Xiangyu Wu and Yifei Hu and Qigen Hu and Xinchen Luo and Lejian Ren and Zixing Zhang and Qianqian Wang and Kuo Cai and Yunfan Wu and Hongtao Cheng and Zexuan Cheng and Lu Ren and Huanjie Wang and Yi Su and Ruiming Tang and Kun Gai and Guorui Zhou},
      year={2025},
      eprint={2510.11639},
      archivePrefix={arXiv},
      primaryClass={cs.IR},
      url={https://arxiv.org/abs/2510.11639}, 
}

@misc{hong2025generativereasoningrecommendationllms,
      title={Generative Reasoning Recommendation via {LLMs}}, 
      author={Minjie Hong and Zetong Zhou and Zirun Guo and Ziang Zhang and Ruofan Hu and Weinan Gan and Jieming Zhu and Zhou Zhao},
      year={2025},
      eprint={2510.20815},
      archivePrefix={arXiv},
      primaryClass={cs.IR},
      url={https://arxiv.org/abs/2510.20815}, 
}

@misc{he2026reasoningsemanticidsenhances,
      title={Reasoning over Semantic {IDs} Enhances Generative Recommendation}, 
      author={Yingzhi He and Yan Sun and Junfei Tan and Yuxin Chen and Xiaoyu Kong and Chunxu Shen and Xiang Wang and An Zhang and Tat-Seng Chua},
      year={2026},
      eprint={2603.23183},
      archivePrefix={arXiv},
      primaryClass={cs.IR},
      url={https://arxiv.org/abs/2603.23183}, 
}

@misc{liang2026generativereasoningreranker,
      title={Generative Reasoning Re-ranker}, 
      author={Mingfu Liang and Yufei Li and Jay Xu and Kavosh Asadi and Xi Liu and Shuo Gu and Kaushik Rangadurai and Frank Shyu and Shuaiwen Wang and Song Yang and Zhijing Li and Jiang Liu and Mengying Sun and Fei Tian and Xiaohan Wei and Chonglin Sun and Jacob Tao and Shike Mei and Wenlin Chen and Santanu Kolay and Sandeep Pandey and Hamed Firooz and Luke Simon},
      year={2026},
      eprint={2602.07774},
      archivePrefix={arXiv},
      primaryClass={cs.IR},
      url={https://arxiv.org/abs/2602.07774}, 
}

@misc{li2025leadremultifacetedknowledgeenhanced,
      title={LEADRE: Multi-Faceted Knowledge Enhanced LLM Empowered Display Advertisement Recommender System}, 
      author={Fengxin Li and Yi Li and Yue Liu and Chao Zhou and Yuan Wang and Xiaoxiang Deng and Wei Xue and Dapeng Liu and Lei Xiao and Haijie Gu and Jie Jiang and Hongyan Liu and Biao Qin and Jun He},
      year={2025},
      eprint={2411.13789},
      archivePrefix={arXiv},
      primaryClass={cs.IR},
      url={https://arxiv.org/abs/2411.13789}, 
}

@misc{damico2026deployingsemanticidbasedgenerative,
      title={Deploying Semantic {ID}-based Generative Retrieval for Large-Scale Podcast Discovery at Spotify}, 
      author={Edoardo D'Amico and Marco De Nadai and Praveen Chandar and Divita Vohra and Shawn Lin and Max Lefarov and Paul Gigioli and Gustavo Penha and Ilya Kopysitsky and Ivo Joel Senese and Darren Mei and Francesco Fabbri and Oguz Semerci and Yu Zhao and Vincent Tang and Brian St. Thomas and Alexandra Ranieri and Matthew N. K. Smith and Aaron Bernkopf and Bryan Leung and Ghazal Fazelnia and Mark VanMiddlesworth and Timothy Christopher Heath and Petter Pehrson Skiden and Alice Y. Wang and Doug J. Cole and Andreas Damianou and Maya Hristakeva and Reid Wilbur and Tarun Chillara and Vladan Radosavljevic and Pooja Chitkara and Sainath Adapa and Juan Elenter and Bernd Huber and Jacqueline Wood and Saaketh Vedantam and Jan Stypka and Sandeep Ghael and Martin D. Gould and David Murgatroyd and Yves Raimond and Mounia Lalmas and Paul N. Bennett},
      year={2026},
      eprint={2603.17540},
      archivePrefix={arXiv},
      primaryClass={cs.IR},
      url={https://arxiv.org/abs/2603.17540}, 
}

@article{denadai2026unified,
  title = {A Unified Language Model for Large Scale Search, Recommendation, and Reasoning},
  author = {De Nadai, Marco and D'Amico, Edoardo and Lefarov, Max and Tamborrino, Alexandre and Vohra, Divita and VanMiddlesworth, Mark and Lin, Shawn and Wood, Jacqueline and Stypka, Jan and Klyce, Eliza and Dai, Keshi and Heath, Timothy Christopher and Gould, Martin D. and Raimond, Yves and Ghael, Sandeep and Jebara, Tony and Damianou, Andreas and Radosavljevic, Vladan and Bennett, Paul N. and Lalmas, Mounia and Chandar, Praveen},
  journal = {arXiv preprint arXiv:2603.17533},
  year = {2026}
}

@article{qwen2025qwen3,
  title = {{Qwen3} Technical Report},
  author = {Yang, An and others},
  journal = {arXiv preprint arXiv:2505.09388},
  year = {2025}
}

@article{microsoft2025phi4reasoning,
  title = {{Phi-4-reasoning} Technical Report},
  author = {Abdin, Marah and Agarwal, Sahaj and Awadallah, Ahmed and Balachandran, Vidhisha and Behl, Harkirat and Chen, Lingjiao and de Rosa, Gustavo and Gunasekar, Suriya and Javaheripi, Mojan and Joshi, Neel and Kauffmann, Piero and Lara, Yash and Mendes, Caio C{\'e}sar Teodoro and Mitra, Arindam and Nushi, Besmira and Papailiopoulos, Dimitris and Saarikivi, Olli and Shah, Shital and Shrivastava, Vaishnavi and Vineet, Vibhav and Wu, Yue and Yousefi, Safoora and Zheng, Guoqing},
  journal = {arXiv preprint arXiv:2504.21318},
  year = {2025}
}

@misc{anthropic2025extendedthinking,
  title = {{Claude 3.7 Sonnet and Claude Code}},
  author = {{Anthropic}},
  year = {2025},
  howpublished = {\url{https://www.anthropic.com/news/claude-3-7-sonnet}},
  note = {Accessed: 2026-04-29}
}

@article{hao2025oxygenrec,
  title = {{OxygenREC}: An Instruction-Following Generative Framework for E-commerce Recommendation},
  author = {Hao, Xuegang and Zhang, Ming and Li, Alex and Qian, Xiangyu and Ma, Zhi and Zang, Yanlong and Yang, Shijie and Han, Zhongxuan and Ma, Xiaolong and Liu, Jinguang and Li, Zhen and Jiang, Zhida and Wang, Shusheng and Tang, Ning and Qiao, Yanchen and Yang, Chenxiang and Sun, Chen and Yuan, Jincheng and Peng, Chunhua and Hu, Heng and Yang, Peijun and Yuan, Baopeng and Qiu, Caiyun and Xing, Zhaolong and Yuan, Haofei and Zhang, Haipeng and Guo, Yuzhang and Ding, Weijie and Gao, Jiahua and Huang, Hao and Chen, Zhen and Liu, Tongxuan and Gong, Pinghua},
  journal = {arXiv preprint arXiv:2512.22386},
  year = {2025}
}

@article{nguyen2026amem4rec,
  title = {{AMEM4Rec}: Leveraging Cross-User Similarity for Memory Evolution in Agentic {LLM} Recommenders},
  author = {Nguyen, Minh-Duc and Kieu, Hai-Dang and Le, Dung D.},
  journal = {arXiv preprint arXiv:2602.08837},
  year = {2026}
}

@article{yi2025recgptv2,
  title = {{RecGPT-V2} Technical Report},
  author = {Yi, Chao and Chen, Dian and Guo, Gaoyang and Tang, Jiakai and Wu, Jian and Yu, Jing and Zhang, Mao and Chen, Wen and Yang, Wenjun and Luo, Yujie and Jiang, Yuning and Gao, Zhujin and Zheng, Bo and Cao, Binbin and Wu, Changfa and Wang, Dixuan and Wu, Han and Hu, Haoyi and Zhu, Kewei and Tian, Lang and Yang, Lin and Huang, Qiqi and Yang, Siqi and Su, Wenbo and He, Xiaoxiao and Tong, Xin and Chen, Xu and Xi, Xunke and Huang, Xiaowei and Wu, Yaxuan and Yang, Yeqiu and Hu, Yi and Yuan, Yujin and Yan, Yuliang and Zhou, Zile},
  journal = {arXiv preprint arXiv:2512.14503},
  year = {2025}
}

@article{tang2025recbot,
  title = {Interactive Recommendation Agent with Active User Commands},
  author = {Tang, Jiakai and Luo, Yujie and Xi, Xunke and Sun, Fei and Feng, Xueyang and Dai, Sunhao and Yi, Chao and Chen, Dian and Gao, Zhujin and Li, Yang and Chen, Xu and Chen, Wen and Wu, Jian and Jiang, Yuning and Zheng, Bo},
  journal = {arXiv preprint arXiv:2509.21317},
  year = {2025}
}

@article{doh2025talkplay,
  title = {{TalkPlay-Tools}: Conversational Music Recommendation with {LLM} Tool Calling},
  author = {Doh, Seungheon and Choi, Keunwoo and Nam, Juhan},
  journal = {arXiv preprint arXiv:2510.01698},
  year = {2025}
}

@misc{wei2023chain,
      title={Chain-of-Thought Prompting Elicits Reasoning in Large Language Models}, 
      author={Jason Wei and Xuezhi Wang and Dale Schuurmans and Maarten Bosma and Brian Ichter and Fei Xia and Ed Chi and Quoc Le and Denny Zhou},
      year={2023},
      eprint={2201.11903},
      archivePrefix={arXiv},
      primaryClass={cs.CL},
      url={https://arxiv.org/abs/2201.11903}, 
}

@misc{grpo,
      title={DeepSeekMath: Pushing the Limits of Mathematical Reasoning in Open Language Models}, 
      author={Zhihong Shao and Peiyi Wang and Qihao Zhu and Runxin Xu and Junxiao Song and Xiao Bi and Haowei Zhang and Mingchuan Zhang and Y. K. Li and Y. Wu and Daya Guo},
      year={2024},
      eprint={2402.03300},
      archivePrefix={arXiv},
      primaryClass={cs.CL},
      url={https://arxiv.org/abs/2402.03300}, 
}

@misc{mcauley2015imagebasedrecommendationsstylessubstitutes,
      title={Image-based Recommendations on Styles and Substitutes}, 
      author={Julian McAuley and Christopher Targett and Qinfeng Shi and Anton van den Hengel},
      year={2015},
      eprint={1506.04757},
      archivePrefix={arXiv},
      primaryClass={cs.CV},
      url={https://arxiv.org/abs/1506.04757}, 
}

@misc{ma2019hierarchical,
      title={Hierarchical Gating Networks for Sequential Recommendation}, 
      author={Chen Ma and Peng Kang and Xue Liu},
      year={2019},
      eprint={1906.09217},
      archivePrefix={arXiv},
      primaryClass={cs.IR},
      url={https://arxiv.org/abs/1906.09217}, 
}

@misc{hidasi2016session,
      title={Session-based Recommendations with Recurrent Neural Networks}, 
      author={Balázs Hidasi and Alexandros Karatzoglou and Linas Baltrunas and Domonkos Tikk},
      year={2016},
      eprint={1511.06939},
      archivePrefix={arXiv},
      primaryClass={cs.LG},
      url={https://arxiv.org/abs/1511.06939}, 
}

@misc{rajput2023tiger,
      title={Recommender Systems with Generative Retrieval}, 
      author={Shashank Rajput and Nikhil Mehta and Anima Singh and Raghunandan H. Keshavan and Trung Vu and Lukasz Heldt and Lichan Hong and Yi Tay and Vinh Q. Tran and Jonah Samost and Maciej Kula and Ed H. Chi and Maheswaran Sathiamoorthy},
      year={2023},
      eprint={2305.05065},
      archivePrefix={arXiv},
      primaryClass={cs.IR},
      url={https://arxiv.org/abs/2305.05065}, 
}

@misc{kang2018self,
      title={Self-Attentive Sequential Recommendation}, 
      author={Wang-Cheng Kang and Julian McAuley},
      year={2018},
      eprint={1808.09781},
      archivePrefix={arXiv},
      primaryClass={cs.IR},
      url={https://arxiv.org/abs/1808.09781}, 
}

@misc{zhai2024hstu,
      title={Actions Speak Louder than Words: Trillion-Parameter Sequential Transducers for Generative Recommendations}, 
      author={Jiaqi Zhai and Lucy Liao and Xing Liu and Yueming Wang and Rui Li and Xuan Cao and Leon Gao and Zhaojie Gong and Fangda Gu and Michael He and Yinghai Lu and Yu Shi},
      year={2024},
      eprint={2402.17152},
      archivePrefix={arXiv},
      primaryClass={cs.LG},
      url={https://arxiv.org/abs/2402.17152}, 
}

@misc{bert,
      title={BERT: Pre-training of Deep Bidirectional Transformers for Language Understanding}, 
      author={Jacob Devlin and Ming-Wei Chang and Kenton Lee and Kristina Toutanova},
      year={2019},
      eprint={1810.04805},
      archivePrefix={arXiv},
      primaryClass={cs.CL},
      url={https://arxiv.org/abs/1810.04805}, 
}

@inproceedings{zhou2018din,
  author    = {Zhou, Guorui and Song, Chengru and Zhu, Xiaoqiang and Fan, Ying and Zhu, Han and Ma, Xiao and Yan, Yanghui and Jin, Junqi and Li, Han and Gai, Kun},
  title     = {Deep Interest Network for Click-Through Rate Prediction},
  booktitle = {Proceedings of the 24th ACM SIGKDD International Conference on Knowledge Discovery and Data Mining},
  series    = {KDD '18},
  pages     = {1059--1068},
  year      = {2018},
  publisher = {Association for Computing Machinery},
  address   = {New York, NY, USA},
  location  = {London, United Kingdom},
  doi       = {10.1145/3219819.3219823},
  url       = {https://doi.org/10.1145/3219819.3219823}
}

\newpage
\appendix

\section{Detailed Reward Design for GRPO Training}
\label{app:reward}

We describe the reward function used to train the slow reasoning model \(\mathcal{M}_{\text{slow}}\). 
\begin{equation}
R_{\text{slow}}
= \lambda_{\text{think}} r_{\text{think}}
+ \lambda_{\text{sid}} r_{\text{sid}}
+ \lambda_{\text{hit}} r_{\text{hit}}.
\end{equation}
\subsection{Reasoning Presence (\(r_{\text{think}}\))}
The model must output a non‑empty reasoning block delimited by \texttt{<think>} and \texttt{</think>} tags, with total character length at least 20.

\begin{equation}
r_{\text{think}} = 
\begin{cases}
+1 & \text{if } \texttt{<think>...</think>} \text{ exists and length} \ge 20,\\
-1 & \text{otherwise}.
\end{cases}
\end{equation}

\subsection{SID Format Correctness (\(r_{\text{sid}}\))}
The expected strict format is:
\texttt{<|sid\_begin|><s\_a\_\#><s\_b\_\#><s\_c\_\#><|sid\_end|>},
where each \(\#\) is a code integer. A soft match accepts any pattern that resembles a valid SID.

\begin{equation}
r_{\text{sid}} = 
\begin{cases}
+1 & \text{strict format matched},\\
+0.2 & \text{soft SID pattern matched},\\
-1 & \text{no valid SID found}.
\end{cases}
\end{equation}

\subsection{Hierarchical Hit Reward (\(r_{\text{hit}}\))}
We parse both the ground‑truth and predicted SIDs into three levels \((a,b,c)\). The reward is based on the longest matching prefix:

\begin{equation}
r_{\text{hit}} = 
\begin{cases}
0 & \text{if no prefix matches},\\
1.0 & \text{if only } a \text{ matches},\\
2.0 & \text{if } a,b \text{ match},\\
5.0 & \text{if } a,b,c \text{ match}.\\
\end{cases}
\end{equation}

This hierarchical shaping provides dense learning signals even when perfect prediction is not yet achieved, while still strongly incentivizing completely correct SID outputs.

\section{Pseudocode for Planner Agent Training}
\label{app:agent_training}
\begin{algorithm}[h]
\caption{Planner Agent Training}
\label{alg:main}
\begin{algorithmic}[1]
\STATE \textbf{Input:} User sequences $\mathcal{U}$, ground truth items, base models $\mathcal{M}_{\text{fast}}, \mathcal{M}_{\text{slow}}$, ranking model $\mathcal{M}_{\text{rank}}$.
\STATE Stage 1: supervised warm‑up
\FOR{each user $u \in \mathcal{U}$}
    \STATE $candidates_{50} \gets \mathcal{M}_{\text{fast}}(u, k=50)$
    \STATE Determine path (1/2/3) based on hit position of ground truth.
    \STATE With 20\% probability, flip path 1 to 2 or 3.
    \STATE Record the required tool call sequence as label.
\ENDFOR
\STATE Train $\mathcal{M}_{\text{agent}}$ via cross‑entropy on tool call sequences.
\STATE Stage 2: Agentic RL
\STATE Freeze $\mathcal{M}_{\text{fast}}, \mathcal{M}_{\text{slow}}, \mathcal{M}_{\text{rank}}$.
\STATE Optimize $\mathcal{M}_{\text{agent}}$ with GRPO using reward $R_{\text{total}}$.
\RETURN Trained agent $\mathcal{M}_{\text{agent}}$.
\end{algorithmic}
\end{algorithm}

\section{Baselines}
\label{app:baseline}
\subsection{Traditional Sequential Recommenders.}
\begin{itemize}[leftmargin=*]
\item \textbf{HGN}~\citep{ma2019hierarchical}: A hierarchical graph neural network that captures user–item interactions at multiple granularities via a gated meta‑path mechanism.
\item \textbf{GRU4Rec}~\citep{hidasi2016session}: A classic RNN‑based model that uses gated recurrent units to model user session sequences for next‑item prediction.
\item \textbf{SASRec}~\citep{kang2018self}: A self‑attention based sequential model that adaptively attends to previous items and has become a widely used strong baseline.
\end{itemize}

\subsection{Generative Recommendation Models.}
\begin{itemize}[leftmargin=*]
\item \textbf{TIGER}~\citep{rajput2023tiger}: A pioneering generative recommender that uses Semantic IDs derived from RQ‑VAE and autoregressively decodes the next item’s SID.
\item \textbf{HSTU}~\citep{zhai2024hstu}: A hierarchical sequential transduction architecture for large-scale recommendation; we use a reproducible configuration adapted to the Amazon benchmark.
\end{itemize}

\subsection{Reasoning‑Enhanced Recommendation Models.}
\begin{itemize}[leftmargin=*]

\item \textbf{OneRec-Think}~\citep{liu2025onerecthinkintextreasoninggenerative}: The most relevant competitor to our work. It introduces itemic alignment via multi‑task pre‑training, reasoning activation through supervised fine‑tuning. However, it applies the same reasoning strategy uniformly to all user histories.
\end{itemize}

\section{A Case Study On Slow Rec Model}
\label{app:case_study_slow_rec_model}
We present a concrete example to illustrate why the thinking model requires hard samples. Consider a user whose historical interactions are dominated by nail art stamping equipment: she has purchased Konad stamping polishes (white, black, clear top coat), a stamp and scraper set, 25 pattern plates from Bundle Monster, along with various OPI polishes, base coats, and cuticle removers. Her next actual purchase is a \textit{40-pack nail art stamping plate bundle}.
When the fast recommendation model or the slow recommendation model trained on full samples encounters this sequence, both collapse to popular, high‑frequency predictions that share surface‑level lexical cues with the history. The fast recommendation model recommends metallic nail polish sets and holographic polishes (e.g., China Glaze Hologram), due to the training distribution. The model fails to recognize that the user has already moved beyond general nail polish into the specialized sub‑task of stamping, where the bottleneck is not color polish but pattern plates.
In contrast, the slow recommendation model (trained exclusively on hard samples) successfully generates a correct recommendation with an interpretable reasoning process. Its internal chain‑of‑thought unfolds as:
\textit{The user has purchased stamping polishes in multiple colors, a stamp and scraper, and a set of 25 plates. In the stamping workflow, after acquiring basic plates, users typically expand their design library by buying additional plate bundles. The next logical item is therefore a larger collection of plates.}

This reasoning leverages knowledge (the stamping process) and the understanding of the user. Such reasoning is unlocked only when the model is forced to generalize from challenging examples where shallow co‑occurrence fails.

\section{Latency and Throughput Analysis}
\label{app:latency}

We report absolute latency and throughput on a single A100 GPU (batch size 1, Beauty test set). The fast recommendation model baseline achieves 0.39 s/sample and 2.56 samples/s. When the planner invokes slow reasoning, latency increases to 2.15 s/sample, adding 1.76 s absolute overhead per sample. The overhead is dominated by chain-of-thought generation (1.60 s) rather than SID decoding (0.16 s). For the full test set, the adaptive planner invokes slow reasoning only on 14.8\% of sequences, keeping average latency near 0.65 s/sample and total inference time under 39 minutes, versus 128.7 minutes for uniform slow reasoning. Uniform fast reasoning would take 23.4 minutes, which is 0.6× the adaptive time, while uniform slow reasoning is 3.3× the adaptive time.
\subsection{Implementation Details}
\label{app:impl_details}

We implement our framework using Qwen3.5-4B\citep{qwen3.5} as the base LLM. All models are trained and evaluated on 8 NVIDIA A100 (80GB) GPUs. Below we detail the configurations for each component.

For each dataset, we follow the leave-one-out strategy: the last interaction for each user is used for testing, the second-last for validation, and all preceding ones for training. We filter out users and items with fewer than 3 interactions to ensure sufficient sequence length \citep{kang2018self}. Item metadata (title and category) are concatenated as the textual description \(t_i\). We generate Semantic IDs with \(L=3\) quantization layers and a codebook size \(K=256\) per layer, resulting in a vocabulary of \(3 \times 256 = 768\) SID tokens. The residual k-means model is trained on the item embeddings from text encoder \citep{bert}.
\subsubsection{Aligned Base Model (\(\mathcal{M}_{\text{align}}\))}
The base LLM (Qwen3.5-4B) is fine-tuned for 3 epochs with a learning rate of \(2\mathrm{e}{-5}\) and a cosine decay scheduler. We freeze all parameters except the embedding matrix of the 768 SID tokens. The batch size is set to 64 sequences, each truncated or padded to a maximum length of 512 tokens. The training objective is the standard cross-entropy loss for next-token prediction.

\subsubsection{Fast Rec Model (\(\mathcal{M}_{\text{fast}}\)) and Ranker}
Starting from \(\mathcal{M}_{\text{align}}\), we train \(\mathcal{M}_{\text{fast}}\) for 5 epochs on the sequential recommendation task. The learning rate is reduced to \(1\mathrm{e}{-5}\) and the batch size is 128. During inference, we use beam search with a width of \(k=50\) and a length penalty of 1.0. The prefix trie for valid SIDs is constructed from the training set's item catalog.

\subsubsection{Slow Rec Model (\(\mathcal{M}_{\text{slow}}\)) with Collaborative Reasoning}
We first generate I2I pairs based on co-occurrence in the same session or sequence. For each pair, we use Qwen3.5-397B-A17B via API to generate a reasoning explanation, resulting in 20,000 final triples after filtering low-quality responses (e.g., length < 30 tokens). The teacher model is prompted with the instruction: ``In collaborative filtering, item \(i\) and item \(j\) are highly correlated. Please explain why users who purchase \(i\) also tend to purchase \(j\)?'' These triples are mixed with the alignment data at a 1:1 ratio to fine-tune \(\mathcal{M}_{\text{align}}\) for 2 epochs with a learning rate of \(1\mathrm{e}{-6}\), producing the base model with collaborative commonsense.

To further train \(\mathcal{M}_{\text{slow}}\), we select training sequences where \(\mathcal{M}_{\text{fast}}\) fails to retrieve the ground truth within top-50 (hard samples). The model is then optimized using GRPO \citep{grpo}. We use a sampling temperature of 0.7, and the policy is updated with a clipping parameter \(\epsilon=0.2\). The reward function is detailed in Appendix~\ref{app:reward}, with weights set to \(\lambda_{\text{hit}}=5.0\) and \(\lambda_{\text{format}}=1.0\). The maximum generation length for the thinking chain is 256 tokens.

\subsubsection{Planner Agent (\(\mathcal{M}_{\text{agent}}\)) Training}
The agent model is initialized from \(\mathcal{M}_{\text{align}}\). For supervised warm-up, we generate pseudo-labels on 100,000 user sequences from the training set, using the path selection logic in Section~\ref{planner_agent} (with probe sizes \(K_1=10\) and \(K_2=50\)). We train \(\mathcal{M}_{\text{agent}}\) for 3 epochs with a learning rate of \(1\mathrm{e}{-5}\) and a batch size of 32. The tool call is formatted as a structured JSON object. For RL fine-tuning, we employ GRPO for another 2 epochs.


\end{document}